\documentclass[aps,reprint,pra,superscriptaddress]{revtex4-2}

\usepackage{amsmath,amssymb,graphicx,color,verbatim,soul}\usepackage{scalerel,stackengine}\usepackage{upgreek}\usepackage{float}\usepackage{bm}\usepackage{hyperref}\hypersetup{colorlinks=true,
urlcolor = blue}

\usepackage{amsmath,amssymb,graphicx,color,verbatim,soul}
\usepackage{scalerel,stackengine}
\usepackage{upgreek}
\usepackage{float}
\usepackage{bm}
\usepackage{hyperref}
\hypersetup{colorlinks=true, urlcolor = blue}
\usepackage{braket}

\begin{document}
\title{Quantum field simulator for dynamics in curved spacetime}

\author{Celia Viermann}
    \affiliation{Kirchhoff-Institut f\"{u}r Physik, Universit\"{a}t Heidelberg, \\
    Im Neuenheimer Feld 227, 69120 Heidelberg, Germany}
\author{Marius Sparn}
    \affiliation{Kirchhoff-Institut f\"{u}r Physik, Universit\"{a}t Heidelberg, \\
    Im Neuenheimer Feld 227, 69120 Heidelberg, Germany}
\author{Nikolas Liebster}
    \affiliation{Kirchhoff-Institut f\"{u}r Physik, Universit\"{a}t Heidelberg, \\
    Im Neuenheimer Feld 227, 69120 Heidelberg, Germany}  
\author{Maurus Hans}
    \affiliation{Kirchhoff-Institut f\"{u}r Physik, Universit\"{a}t Heidelberg, \\
    Im Neuenheimer Feld 227, 69120 Heidelberg, Germany}
\author{Elinor Kath}
    \affiliation{Kirchhoff-Institut f\"{u}r Physik, Universit\"{a}t Heidelberg, \\
    Im Neuenheimer Feld 227, 69120 Heidelberg, Germany}  
\author{Álvaro Parra-López}
    \affiliation{Institut f\"{u}r Theoretische Physik, Universit\"{a}t Heidelberg, \\ Philosophenweg 16, 69120 Heidelberg, Germany}
    \affiliation{Departamento de F\'isica Te\'orica and IPARCOS, Facultad de Ciencias Físicas, Universidad Complutense de Madrid, Ciudad Universitaria, 28040 Madrid, Spain}
\author{Mireia Tolosa-Simeón}
    \affiliation{Institut f\"{u}r Theoretische Physik, Universit\"{a}t Heidelberg, \\ Philosophenweg 16, 69120 Heidelberg, Germany}
\author{Natalia Sánchez-Kuntz}
   \affiliation{Institut f\"{u}r Theoretische Physik, Universit\"{a}t Heidelberg, \\ Philosophenweg 16, 69120 Heidelberg, Germany}
\author{Tobias Haas}
    \affiliation{Institut f\"{u}r Theoretische Physik, Universit\"{a}t Heidelberg, \\ Philosophenweg 16, 69120 Heidelberg, Germany}
    \author{Helmut Strobel}
    \affiliation{Kirchhoff-Institut f\"{u}r Physik, Universit\"{a}t Heidelberg, \\
    Im Neuenheimer Feld 227, 69120 Heidelberg, Germany}
\author{Stefan Floerchinger}
    \affiliation{Institut f\"{u}r Theoretische Physik, Universit\"{a}t Heidelberg, \\ Philosophenweg 16, 69120 Heidelberg, Germany}
    \affiliation{Theoretisch-Physikalisches Institut, Friedrich-Schiller-Universit\"{a}t Jena,\\
    Max-Wien-Platz 1, 07743 Jena, Germany}
    \author{Markus K. Oberthaler}
    \affiliation{Kirchhoff-Institut f\"{u}r Physik, Universit\"{a}t Heidelberg, \\
    Im Neuenheimer Feld 227, 69120 Heidelberg, Germany}

\maketitle
\textbf{ The observed large-scale structure in our Universe is seen as a result of quantum fluctuations amplified by spacetime evolution \cite{Springel2006}. This, and related problems in cosmology, asks for an understanding of the quantum fields of the standard model and dark matter in curved spacetime.
Even the reduced problem of a scalar quantum field in an explicitly time-dependent spacetime metric is a theoretical challenge \cite{Parker1969,Birrell1982,Mukhanov2007} and thus a quantum field simulator can lead to new insights.
Here, we demonstrate such a quantum field simulator in a two-dimensional Bose-Einstein condensate with a configurable trap \cite{SaintJalm2019,Gauthier2021} and adjustable interaction strength to implement this model system. 
We explicitly show the realisation of spacetimes with positive and negative spatial curvature by wave packet propagation and confirm particle pair production in controlled power-law expansion of space. We find quantitative agreement with new analytical predictions for different curvatures in time and space.
This benchmarks and thereby establishes a quantum field simulator of a new class. 
In the future, straightforward upgrades offer the possibility to enter new, so far unexplored, regimes that give further insight into relativistic quantum field dynamics.}

As early as 1980, W.G. Unruh \cite{Unruh1981} pointed out that the motion of sound waves in a convergent fluid flow can be seen in close analogy to the behaviour of a quantum field in a classical gravitational field. It is thus directly connected to quantum field theory in curved spacetime. Many impressive experiments and theoretical studies have followed, building on the imprint of a flow pattern to generate the metric structure of interest \cite{Unruh1995,Garay2000,Visser2002,Novello2002, Barcelo2003c,Fedichev2004,Weinfurtner2007,Schuetzhold2009,Prain2010, Barcelo2011,Jacquet2020}. This ranges from classical \cite{Philbin2008,Weinfurtner2011} to quantum fluids \cite{Carusotto2008,Lahav2010,Steinhauer2014,Eckel2018,MunozDeNova2019,Wittemer2019,Banik2021}. 
Here, we show that even a static but spatially inhomogeneous superfluid can provide a curved spacetime metric for its phononic excitations. Using a Bose-Einstein condensate of ultracold atoms, spatial curvature is adjustable by the proper choice of the density distribution, while time-dependence of the metric is implemented via a change in the inter-atomic interaction.  

In our experiment, we realise a two-dimensional condensate of potassium-39 with configurable density distribution and additional dynamic control of atomic interactions. With that we implement curved metrics for the phononic field of the form (see methods: `Curvature in a metric')
\begin{equation}
\mathrm d s^2 = - \mathrm d t ^2 + a^2 (t)  \left(\frac{\mathrm d u^2}{1- \kappa {u^2}} + u^2 \mathrm d \varphi^2\right).
\label{equ:FLRWmetric}
\end{equation}
This corresponds to the standard cosmological metric of a $2+1$ dimensional homogeneous and isotropic universe, the Friedmann-Lemaître-Robertson-Walker metric (FLRW) in reduced circumference coordinates $(u,\varphi)$. This metric is parametrised by intrinsic and extrinsic curvature: 
 the intrinsic curvature, $\kappa$, is the curvature of the spatial part of the metric, while the extrinsic curvature arises from the time dependence of the scale factor $a(t)$. In our atomic implementation both parameters, curvature and scale factor, can be adjusted independently.

\begin{figure*}
    \centering
    \includegraphics[width = \textwidth]{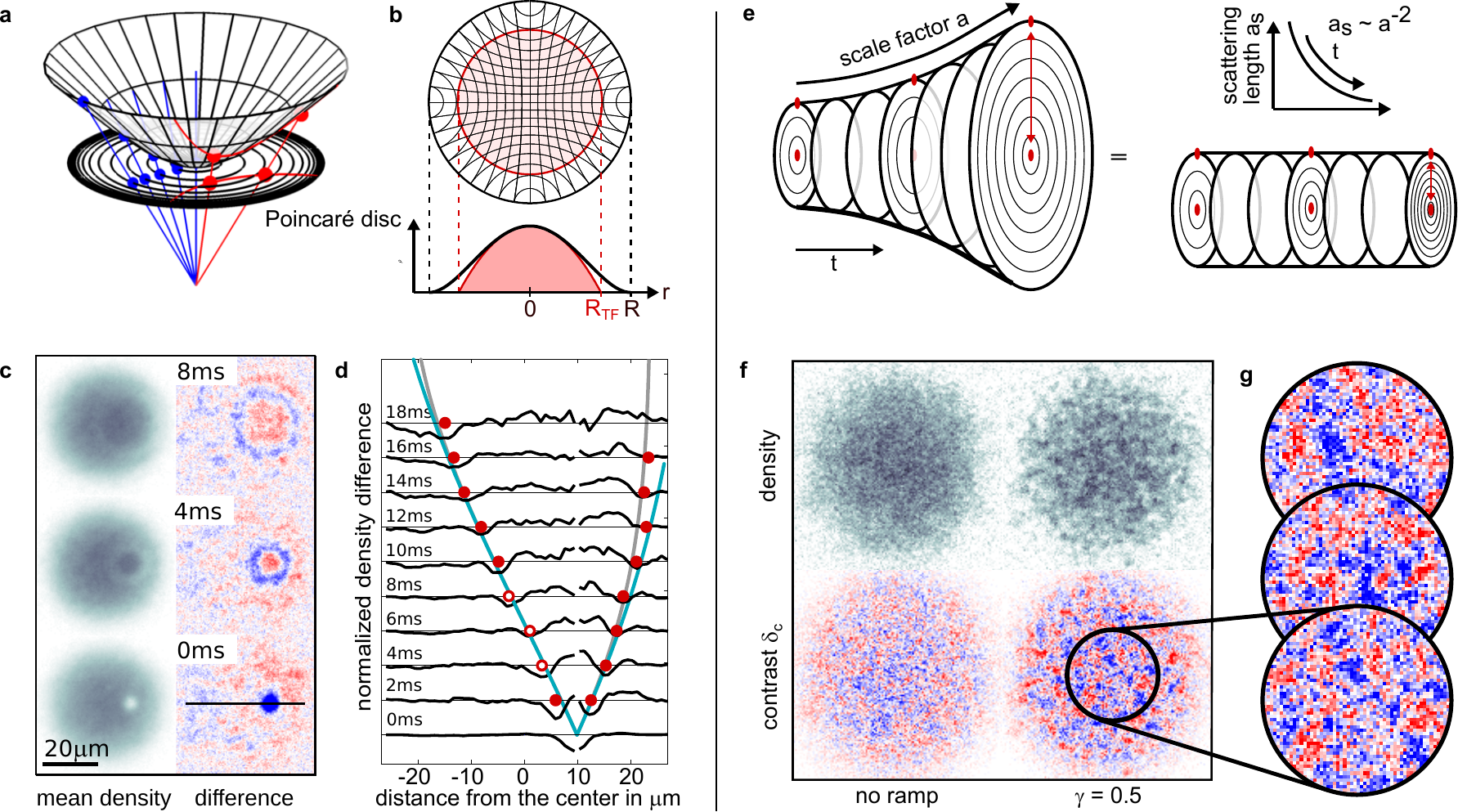}
     \caption{\textsf{\textbf{Curvature in space and time realised in a Bose-Einstein condensate.}
     \textbf{a)} Hyperbolic space with constant negative curvature mapped onto the finite-sized Poincaré disc. \textbf{b)} Realisation of a hyperbolic geometry in an inhomogeneous condensate. The corresponding density profile (black) is approximated by a condensate in a harmonic trap (red).
     \textbf{c)} Propagation of a phononic wave packet averaged over $\sim 100$ realisations (left) and difference to the unperturbed condensate (right).
     \textbf{d)} Quantitative comparison between prediction and experiment for the propagation along the geodesic indicated in the lower right panel of c (black line). The red dots mark the position of the wave packet at each time. The blue line is the theory prediction for the hyperbolic space and the grey line the prediction for the acoustic metric of the parabolic Thomas-Fermi profile.
     \textbf{e)} Illustration of the equivalence between an expanding space and the static BEC with dynamically controlled s-wave scattering length $a_s$.
     \textbf{f)}  Density and density contrast $\delta_c$ of a single realisation before and after a ramp with scale factor $a(t) \propto t^{\gamma}$. The emergence of fluctuations on large scales indicates particle production.  \textbf{g)} Structures are distributed randomly in different realisations. 
    }}
    \label{fig:Fig1}
\end{figure*}

Phonons in the central region of a harmonically trapped Bose-Einstein condensate experience a metric with $\kappa < 0$. In cosmological settings, this is a hyperbolic two-dimensional spatial geometry with a radial coordinate of infinite range, as depicted in Figure \ref{fig:Fig1} \textbf{a}. Through the Poincaré transformation, the infinite hyperbolic space is mapped to a finite disc, perfectly suited to be implemented in finite-size ultracold gases. The Poincaré disc in Figure \ref{fig:Fig1} \textbf{a} shows equidistant lines in this metric, which become more dense towards the edges, illustrating the divergence of distances. 

The exact density profile for the implementation of a hyperbolic metric is shown in Figure \ref{fig:Fig1} \textbf{b} (black line) which is naturally realised in the central region of a harmonically trapped Bose-Einstein condensate (red curve). The corresponding spatial curvature is given by ${\kappa = -2/R_\text{TF}^2}$, with the size of the cloud given by its Thomas-Fermi radius $R_\text{TF}$.

To demonstrate the implementation of hyperbolic space with our condensate, we observe the propagation of a wave packet that follows geodesics of the underlying metric. We prepare the wave packet by focusing a blue-detuned laser beam onto the atomic cloud, which evolves into an expanding wave in the hyperbolic geometry once the laser is switched off. Figure \ref{fig:Fig1} \textbf{c} shows this propagation averaged over $\sim 100$ realisations, as well as the density difference to the unperturbed system. 
For each time-step, the profile of the density is extracted along the geodesic connecting the initial perturbation with the centre of the condensate.

Figure \ref{fig:Fig1} \textbf{d} shows the normalised profiles from which the positions of the minima are extracted (red points). We use the three points marked with open symbols to fit the speed of sound at the centre of the condensate. This measured speed of sound and the extracted Thomas-Fermi radius completely determine the metric and hence set the prediction for the phononic wave packet trajectories (for further details see methods). The solid grey line in Fig. \ref{fig:Fig1} \textbf{d} shows the prediction for the harmonically trapped condensate, and the blue line the prediction for a hyperbolic space. This serves as a quantitative demonstration that a condensate in a harmonic trap approximates a hyperbolic geometry, corresponding to negative curvature. Deviations only occur close to the Thomas-Fermi radius of $25\, \mu$m. 

\begin{figure}
    \centering
    \includegraphics[width=0.48\textwidth]{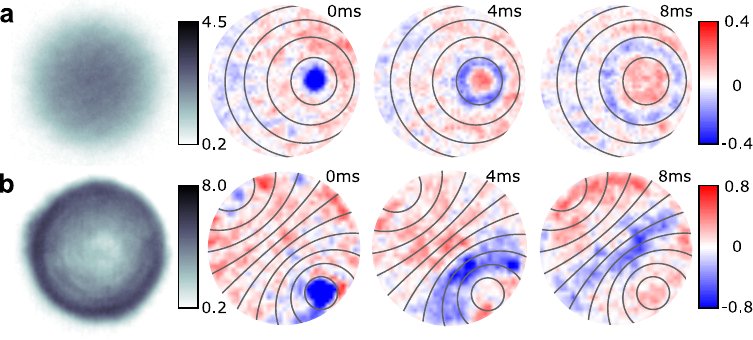}
    \caption{\textsf{\textbf{Configurable density distribution for hyperbolic and spherical geometry.}
    \textbf{a)} Density distribution of the condensate for hyperbolic geometry and expansion of a phononic wave packet depicted in the difference to the unperturbed condensate (blue under-, red over-density). For improved visibility, a two-by-two region of pixels is averaged in density-difference panels.
    \textbf{b)} Density distribution with maximal density at the rim of the two-dimensional condensate realises spherical geometry. The propagation of the wave packet exhibits the expected straightening confirming the successful implementation of the metric with positive curvature. Black lines indicate equidistant lines from the initial perturbation in the hyperbolic and spherical metric, respectively.}}
    \label{fig:FigX}
\end{figure}

This concept can be extended to different curvatures by choosing the appropriate density profile (see methods). For spacetime geometries beyond hyperbolic, we use a digital micromirror device (DMD) \cite{Gauthier2021,SaintJalm2019} to configure arbitrary spatial curvatures.  One such possibility is positive spatial curvature, also known as spherically curved space. Figure \ref{fig:FigX} contrasts the wave packet propagation on background densities corresponding to hyperbolic and spherical metrics. We observe fundamentally different evolution in agreement to the expected dynamics. This confirms that our simulator can be configured to positive or negative curvature.

The implementation of curvature in time, i.e. extrinsic curvature, is depicted in Figure \ref{fig:Fig1} \textbf{e} in two equivalent representations of a spatially flat FLRW metric with increasing scale factor $a(t)$. The distance covered by a signal moving at the causal speed in a unit of time is depicted by the separation between the circles. 
For a constant causal speed, the expansion of space is encoded in the increase of coordinate distance between the two red points, as illustrated on the left side of \ref{fig:Fig1} \textbf{e}. However, this is equivalent to keeping coordinates static (comoving coordinates), while decreasing the causal speed to capture the expansion (right side of \ref{fig:Fig1} \textbf{e}). 

We use this equivalence to implement expanding space in our static BEC by dynamically controlling the s-wave scattering length $a_s(t)$, which relates to the scale factor $a(t)$ via
\begin{equation} \label{scaleFactor_scatLength}
a^2(t) = \sqrt{\frac{m^3}{8 \pi \, \omega_z \, \hbar^3 \,\bar n_0^2}}\, \frac{1}{a_s(t)},
\end{equation}
where $\omega_z = 2\pi \cdot 1.6\, \mathrm{kHz} $ is the trap frequency in the tightly confined direction, $\bar n_0 = 1.3\cdot10^9 \, \mathrm{cm^{-2}}$ is the two-dimensional density at the centre of the trap and $m$ is the atomic mass. 
Control of the scattering length is achieved by changing an external magnetic field in the vicinity of the Feshbach resonance in potassium at $562.2(1.5)\, \mathrm{G}$ \cite{DErrico2007}. In order to keep the density distribution of the condensate constant during changes of the scattering length, we adjust the radial trap frequency accordingly.

In quantum field theory, the expansion of space leads to the phenomenon of particle production \cite{Parker1969,Birrell1982,Mukhanov2007}. Motivated by cosmological expansion scenarios, we perform power-law ramps with the scale factor $a(t) \propto t^\gamma$, where $\gamma = 1$ corresponds to a uniformly expanding universe, $\gamma > 1$ an accelerating universe and $\gamma < 1$ a decelerating universe. In Figure \ref{fig:Fig1} \textbf{f} typical density distributions before and after a specific ramp with $\gamma = 0.5$ (decelerating universe) are shown. After the ramp, the atomic gas features enhanced density fluctuations as expected from particle production. These enhanced fluctuations have also been observed in quench experiments, corresponding to an infinitely fast ramp, performed in atomic \cite{Hung2013,Chen2021} and photonic systems \cite{Steinhauer2021}. We show in the following that the shape of the ramp leads to distinctive features in the emerging quantum state. 

\begin{figure}[t]
    \centering
    \includegraphics[width = 0.48\textwidth]{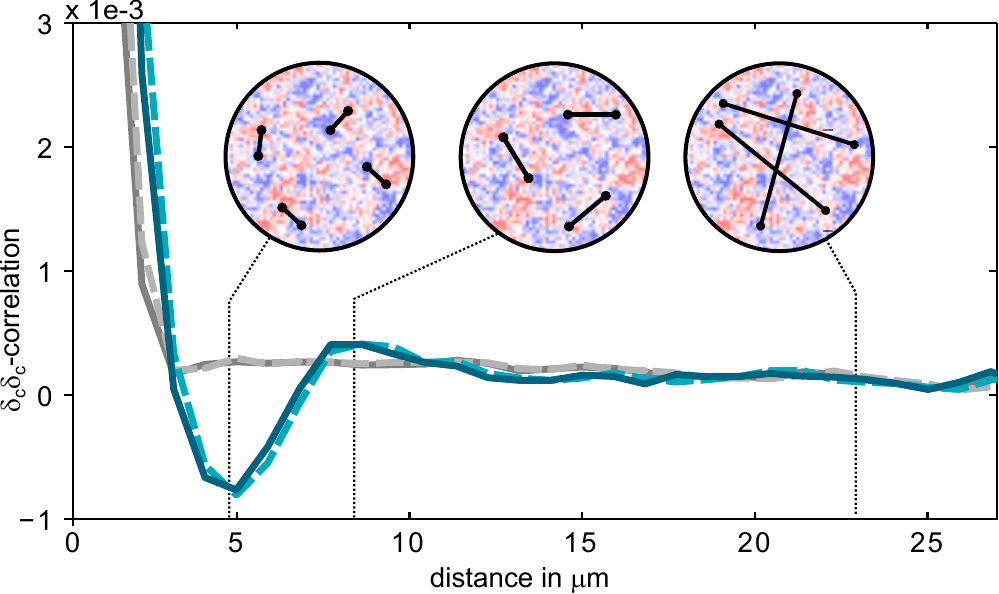}
     \caption{\textsf{\textbf{Correlation function of fluctuations before and after the ramp.} Correlation functions are extracted from the  density contrast $\delta_c$ (see Eq. \eqref{eq:density_contrast}) as an average over all pairs of points with the same distance. The correlation function after a ramp (blue line) shows clear structure in comparison to the correlation function before the ramp (grey line). Distances are evaluated in the flat (solid lines) and in the hyperbolic metric (dashed lines). As expected from the geometry of the Poincare disk in its central region, these two are very close.}}
    \label{fig:Fig2} 
\end{figure}

For a comparison between theory and experiment we consider the experimentally accessible density contrast
\begin{equation}
    \delta_c(x,y) = \sqrt{\frac{  n_0(x,y)}{\bar n_0^3}}\left[n(x,y)-  n_0(x,y)\right]\;, 
    \label{eq:density_contrast}
\end{equation}
with the density distribution of the background condensate $n_0(x,y) = \langle n(x,y)\rangle$, inferred from the mean over all experimental realisations at each pixel position with coordinates $x$ and $y$.
This specific form is chosen such that the density contrast is proportional to the time derivative of a massless, real, relativistic scalar field $\phi$, describing phononic excitations in a curved spacetime governed by the effective action
 \begin{equation} \label{eq:action}
    \Gamma = -\frac{\hbar^2}{2} \int \text{d} t \, \text{d}u  \text{d} \varphi \, \sqrt{g} \, g^{\mu\nu} \partial_\mu \phi \, \partial_\nu \phi\;,
\end{equation}
with the metric $g_{\mu\nu}$ and $\sqrt{g}=\sqrt{-\text{det} (g_{\mu\nu})}$ (see details in \cite{BECPaper2022}).
For expanding spacetimes, this is the minimal model system for cosmological particle pair production. This quantum process is visible as enhanced fluctuations in single experimental realisations (see Figure \ref{fig:Fig1} \textbf{g}), which we quantify with the $\delta_c$-$\delta_c$-correlation function, averaged over many realisations.
The symmetries of the underlying metric imply that the correlation depends only on the distance $L$ measured with the metric in Eq. \eqref{equ:FLRWmetric}  between two points. Thus, we evaluate correlations as an average over all pixel-pairs of given hyperbolic distance in the set of experimental realisations (see Figure \ref{fig:Fig2}).

Figure \ref{fig:Fig2} \textbf{b} shows the correlation functions before (grey) and after (blue) a ramp with $a(t) \propto t^{0.5}$ evaluated in a circular region at the centre of the cloud with diameter $R_\text{TF}$. 
While there are only short-range correlations of the initial condensate over the area of interest, we find a clear anti-correlation signal at a length scale of $5 \, \mu$m after the decelerated expansion, followed by a small correlation peak. We find the same correlation function for an analysis with distances evaluated in the hyperbolic and in the flat metric, as expected for the central region of the Poincaré disc. 

\begin{figure}[t]
    \centering
    \includegraphics[width = 0.48\textwidth]{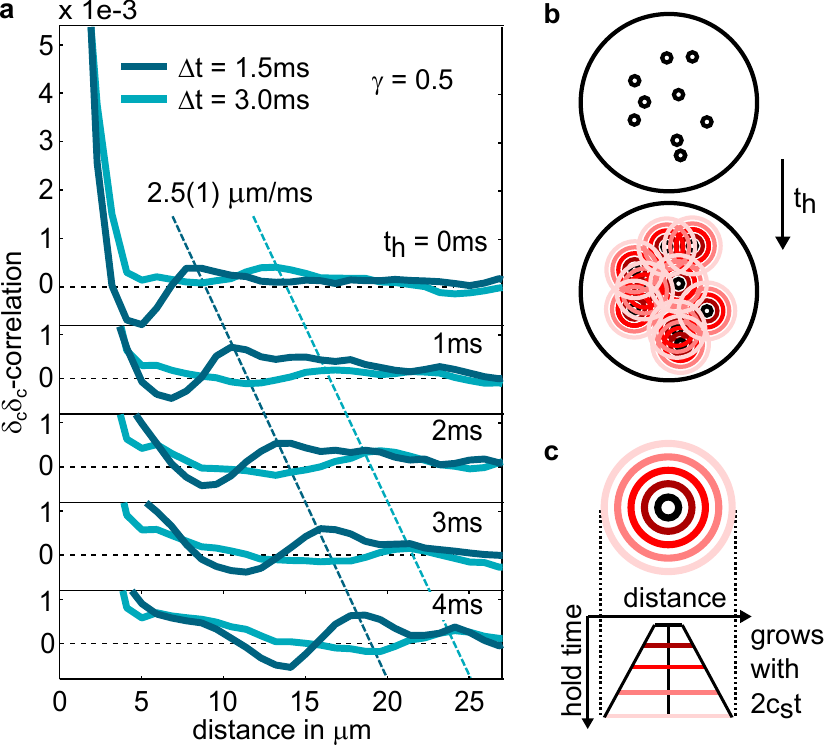}
     \caption{ \textsf{\textbf{Propagation of correlations after expansion.} 
     \textbf{a)} The characteristic features in the correlation functions propagate to larger distances, shown for two different ramp durations. The speed of propagation is twice the speed of sound in both cases. These are Sakharov oscillations in real space. 
     \textbf{b)} Localised excitations at random initial positions emit waves that propagate at the speed of sound (red circles).
     \textbf{c)} A correlation is expected at the diameter of the sound cone, which grows with twice the speed of sound.
     }}
    \label{fig:Fig3}
\end{figure}

An advantage of the atomic system is that the time evolution of the quantum field after the expansion can be directly accessed, in contrast to cosmological settings. 
For this, we let the system evolve at constant scattering length $50 a_B$ for a hold time $t_h$ after the ramp. Figure \ref{fig:Fig3} \textbf{a} shows the correlation functions at different hold times for $\gamma = 0.5$ and two different ramp durations.
For the slower ramp, the characteristic correlation feature appears at a larger distance and is less pronounced. Correlations move to larger distances with a constant speed of $v = 2.5(1) \, \mu \mathrm{m}/\mathrm{ms}$, for both ramp durations. The speed agrees with twice the speed of sound in the centre of the condensate predicted from the density of the three-dimensional GPE ground state. 
The propagation of the correlation is the result of the time evolution of the spontaneously created excitations: initial excitations spread radially outwards with each wavefront moving at the speed of sound. Interference of the many waves leads to a pattern that appears to be random (see Figure \ref{fig:Fig3} \textbf{b}), but the underlying structure shown in \textbf{c} is preserved. A correlation peak is expected at the diameter of the circular wavefront, which increases with twice the speed of sound. 

\begin{figure*}[t!]
    \centering
    \includegraphics[width = \textwidth]{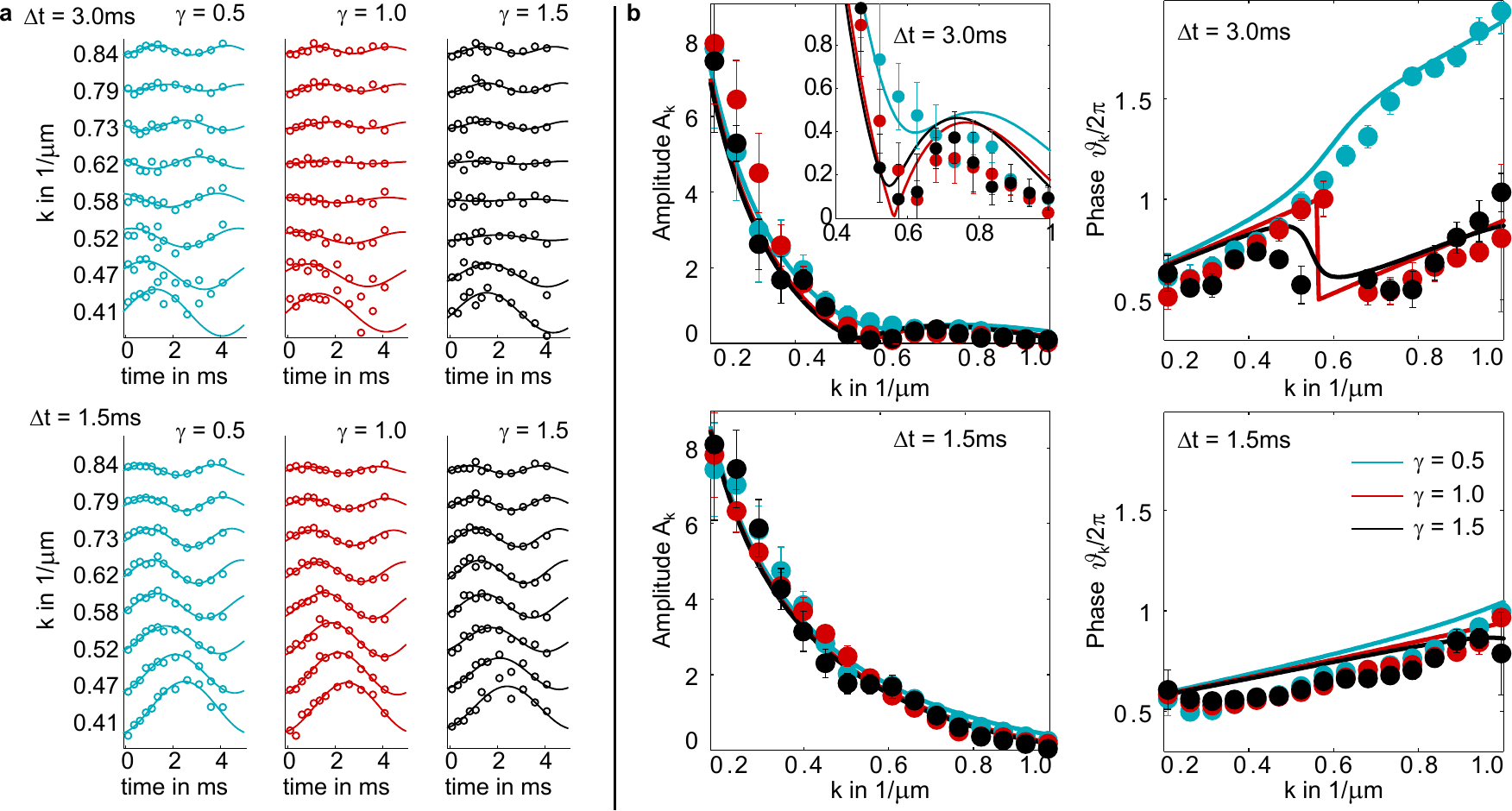}
    \caption{\textsf{\textbf{Expansion histories extracted by heterodyne detection.}
    \textbf{a)} Sakharov oscillations of different $k$-modes  for three different $\gamma$ corresponding to an decelerated (blue), uniformly expanding (red) and accelerated (black) universe together with cosine fits. 
    \textbf{b)} Benchmarking of the quantum field simulator with time dependent correlations.  
    Initial phase and amplitude of the heterodyne signal show quantitative agreement with analytical theory predictions \cite{BECPaper2022} (solid lines). Of special interest is the dramatic signal of the initial phase for the $3\, \mathrm{ms}$ ramp. This is a feature of the produced quantum state, robust against thermal fluctuations in the initial state. 
    }}
    \label{fig:Fig4}
\end{figure*}

Remarkably, the time evolution contains information about the prior expansion history. This is not apparent in the real space correlations, but becomes clear in the evolution of individual momentum modes. 
Since particles are only produced during the ramp, the populations do not change further during the hold time. However, each phononic excitation accumulates phase according to its energy. Experimentally, we observe the phase evolution by the interference between the modes and the background condensate, analogous to heterodyne detection in quantum optics \cite{Chen2021,Gross2011}.

To analyse the resulting oscillations, also known as Sakharov oscillations \cite{Sakharov1966,Grishchuk2012,Hung2013}, we consider the spectral decomposition of the density-contrast correlations. Since the correlations depend only on distances and do not contain angular information, we use the Hankel transformation of zero order, which is the appropriate decomposition for a spatially flat metric. 
The time evolution of specific $k$-modes of the Hankel transform are shown in Figure \ref{fig:Fig4} \textbf{a} for ramps with three different power-laws (colours) and two different ramp speeds. They reveal oscillations with twice the frequency of the phononic dispersion relation of the corresponding $k$. The solid lines are functional fits to the data of the form ${f_k(t_h) = A_k \cdot \cos(2 \omega_k t_h + \vartheta_k) + \text{const}}$. Here, for $\omega_k = c_s k$ we insert the speed of sound previously extracted from the evolution of the correlation functions and $A_k$ and $\vartheta_k$ are fit parameters. Figure \ref{fig:Fig4} \textbf{b} shows amplitude and phase of the oscillations for the two different ramp speeds. The error bars are $1$-s.d. errors from the fit. Here, the solid lines are theory predictions derived by analytically solving mode evolution equations for a free, massless, relativistic scalar field in a $(2+1)$-dimensional expanding spacetime \cite{BECPaper2022}. 

The prediction uses a final sound speed of ${c_s = 1.2\, \mathrm{\mu m/ms}}$ at a scattering length of $50 \mathrm{a_\text{B}}$ and an initial scattering length of $350 \mathrm{a_\text{B}}$. These parameters are within the error bounds of the independently extracted sound speed and the position of the Feshbach resonance. It also uses an initial excitation spectrum which corresponds to a temperature of $T=40\; \mathrm{n K}$. Experimentally, the temperature of the gas in the harmonic trap is independently determined to be $T= 60(10)\; \mathrm{n K }$ using the in-situ density distribution \cite{Giorgini1996}. We ascribe the deviation between experimental and theoretical temperatures to the fact that thermal excitations are expelled from the centre of a harmonically trapped condensate, and thus the initial fluctuations are captured by a lower effective temperature.

The quantitative agreement between analytical theory and the observations of amplitude and phase confirms that we indeed observe the predicted particle production caused by the expanding metric.  
Specifically, for $\gamma = 1.0$ (uniform expansion) and $\gamma =1.5$ (accelerated expansion) we confirm the existence of a phase jump which is not present for $\gamma =0.5$ (decelerated). Theoretically we find that this feature is independent of temperature and thus an ideal indicator for the expansion history of the metric.

Our results confirm the first successful implementation of a foundational model of relativistic quantum field theory - a relativistic scalar quantum field in curved spacetime. With the unique capabilities of the experimental system, fundamental questions in quantum physics can be investigated, such as entanglement in time-evolving curved space  \cite{Berges2018a,Robertson2017,Kunkel2022}, the connection of general event horizons \cite{Gibbons1977}, thermodynamics, and general relativity \cite{Jacobson1995,Jacobson2016}.
Extending the experimental platform to multi-species atomic systems allows to realise unorthodox versions of spacetime geometry \cite{Fischer2004b, SchmidtMay2016}.

\section{Acknowledgements}
The authors would like to thank Simon Brunner and Finn Schmutte for discussions. This work is supported by the Deutsche Forschungsgemeinschaft (DFG, German Research Foundation) under Germany's Excellence Strategy EXC 2181/1 - 390900948 (the Heidelberg STRUCTURES Excellence Cluster) and under SFB 1225 ISOQUANT - 273811115, as well as FL 736/3-1. NL acknowledges support by the Studienstiftung des Deutschen Volkes. NSK is supported by the Deutscher Akademischer Austauschdienst (DAAD, German Academic Exchange Service) under the Länderbezogenes Kooperationsprogramm mit Mexiko: CONACYT Promotion, 2018 (57437340). APL is supported by the MIU (Spain) fellowship FPU20/05603 and the MICINN (Spain) project PID2019-107394GB-I00 (AEI/FEDER,UE).

\section{Author contributions}
The experimental concept was developed in discussion among all authors. M.H., E.K., N.L., M.K.O., M.S., H.S., C.V. controlled the experimental apparatus, discussed the measurement results and analysed the data. S.F., T.H., N.S-K, A.P-L., M.T-S. elaborated the theoretical framework. All authors contributed to the discussion of the results and the writing of the manuscript.
\newpage

\section{Methods - typically do not exceed 3000 words}
\textbf{Experimental system}\\

We prepare a two-dimensional potassium-$39$ Bose-Einstein condensate of approximately $23,000$ atoms in the state that corresponds to $F=1$, $m_F = -1$ at low magnetic field. The atoms are levitated against gravity by a magnetic field gradient. A two-dimensional geometry is achieved by a blue-detuned lattice at $532\, \mathrm {nm}$ with a spacing of $5 \, \mathrm{\mu m}$ leading to a strong confinement with a trap frequency of $\omega_z = 2\pi \cdot 1.6\,\mathrm{kHz}$. Radial trapping is provided by a red-detuned Gaussian beam at $1064\, \mathrm{nm}$ in the case of hyperbolic geometry and a beam shaped by a digital micromirror device in direct imaging configuration for the spherical geometry. The DMD light is blue-detunded at $532\, \mathrm{nm}$. The radial trapping frequency is dynamically adjusted between $23\, \mathrm{Hz}$ and $7\,\mathrm{Hz}$ for the realisation of hyperbolic geometry and metric expansion. The Thomas-Fermi radius of the condensate is typically $25\, \mathrm{\mu m}$ for the curvature measurements and $30\, \mathrm{\mu m}$ for the expansion measurements. The scattering length is adjusted utilising the Feshbach resonance at $562.2(1.5) \,\mathrm G$ \cite{DErrico2007}. For wave packet propagation the scattering length is set to $100 a_\text{B}$ for the hyperbolic case and to $200 a_\text{B}$ for the spherical case. In the expansion experiments the ramps are performed from $400 a_\text{B}$ to $50 a_\text{B}$. During the ramp, the trap frequency is adjusted such that the shape of the condensate remains static.
The density distribution of the atoms is detected with absorption imaging using a two-frequency scheme described in \cite{Hans2021}, which allows an approximately closed scattering cycle at the magnetic fields around the Feshbach resonance. Imaging resolution is $1\, \mathrm{\mu m}$.
\newline
\vspace{0.3mm}

\textbf{Curvature in a metric}\\

The acoustic metric in a Bose-Einstein condensate \cite{Volovik2009,Weinfurtner2007,Bilic2013,BECPaper2022} is given by
\begin{equation}
\mathrm d s^2= - \mathrm d t ^2 +  \frac{1}{c_s^2} \left(\mathrm d r^2 + r^2 \mathrm d \varphi^2\right),
\end{equation}
with $c_s$ the time and space dependent speed of sound. On a two-dimensional condensate with azimuthal symmetry, a hyperbolic geometry with constant curvature is achieved for the radial density profile ${n_0(r) = \bar n_0[1-r^2 /R^2]^2}$ with $\bar n_0$ the density at the centre and $R$ the maximal radius (radius of the Poincaré disc). This yields the speed of sound
\begin{align}
    c_s^2 = \frac{\lambda(t) n_0(r)}{m},
\end{align}
with $\lambda(t)= \sqrt{ 8\pi \omega_z \hbar^3 / m} \, a_s(t)$ the two-dimensional coupling strength depending on the s-wave scattering length $a_s(t)$. This brings the metric into the form 
\begin{equation}
\mathrm d s^2 = - \mathrm d t ^2 + a^2 (t) \left(1- \frac{r^2}{R^2}\right)^{-2} \left(\mathrm d r^2 + r^2 \mathrm d \varphi^2\right) \;.
\end{equation}
The FLRW metric in Eq.\ \eqref{equ:FLRWmetric} is recovered by a coordinate transformation between the radial lab coordinate $r$ and reduced circumference coordinates
\begin{align}
    u(r) = \frac{r}{1 - \frac{r^2}{R^2}}.
    \label{eq:CoordinateTransformation}
\end{align}

A condensate in a harmonic trap has a parabolic density profile with Thomas-Fermi radius $R_\text{TF}$, which approximates the above density profile around the centre of the condensate for $R = \sqrt 2 R_\text{TF}$.

The propagation of a wave packet proceeds along a geodesic. Since phonons propagate with the speed of sound, their trajectories satisfy $\mathrm d s ^2 = 0$ and for the distance $L$ between two points can be measured with the metric in Eq. \eqref{equ:FLRWmetric}
\begin{align} \nonumber
    L(u,\varphi,u',\varphi')&= \frac{1}{\sqrt{|\kappa|}}\cosh^{-1} \big[\sqrt{|\kappa|u^2+1 }\sqrt{|\kappa|u'^2+1} \\
  & -  |\kappa| u u' \cos(\varphi-\varphi') \big].
  \label{equ:hyperbolic_distance}
\end{align}
For the trajectory prediction in Figure \ref{fig:Fig1} \textbf{d}, we use Eq.~\eqref{equ:hyperbolic_distance} with $\varphi=\varphi^\prime$, divided by the speed of sound at the centre of the condensate, and Eq.\ \eqref{eq:CoordinateTransformation}.

The free parameters, speed of sound and Thomas-Fermi radius, are determined to be $R_\text{TF} = 25\,\mathrm{\mu m}$ and $c_s = 1.5\,\mathrm{\mu m/ms}$. The former is extracted from the density profile and the latter is the slope fitted to the three open symbols in Figure \ref{fig:Fig1} \textbf{d}. This is consistent with the speed of sound estimated from GPE simulations of the ground state. 

The experimental density profiles in Figure \ref{fig:Fig1} \textbf{d} are obtained along the black line in the lower right panel of Figure \ref{fig:Fig1} \textbf{c}. To increase signal to noise, we take an angular average over $\pm 10^\circ$ around that line. We determine the position of the wave packet to both sides of the initial position by a parabolic fit around the minimum value of the profile.

An analogous calculation can be done for a positive spatial curvature implemented by the density profile
$ n_0(r) = \bar n_0 [1+ r^2 /R^2]^2$ \cite{BECPaper2022}. 
\newline
\vspace{0.3mm}

\textbf{Extraction of the correlation function}\\

The correlation functions are extracted from the density contrast in Eq. \eqref{eq:density_contrast} within the central region of the condensate up to half the Thomas-Fermi radius. 
For the density contrast, we first compute the distribution of the background condensate as an average over all realisations post-selected to be within $10\%$ of the mean atom number. The background is normalised to the atom number of each realisation. To increase signal to noise, four pixels are averaged, corresponding to the optical resolution. 
\newline
\vspace{0.3mm}

The spectrum is derived from the correlation function by a one-dimensional Hankel transform 
\begin{align}\label{eq:Sk}
S_k  =   \frac{\bar n_0\, m}{\hbar\, a(t_\text{f})}   \frac{1}{k} \int \text{d}L\, L\,J_0(kL)\,\left<\delta_c\delta_c\right>(L),
\end{align}
which is a decomposition into Bessel functions of the first kind $J_0(kL)$. Here, the scale factor $a(t_\text{f})$ is calculated with the scattering length at the end of the ramp $a_s(t_\text{f})$ \eqref{scaleFactor_scatLength}. The transformation in Eq. \eqref{eq:Sk} is appropriate here due to statistical homogeneity and isotropy.

From theory, the power spectrum of fluctuations for non-zero initial temperatures is predicted to be of the form
\begin{equation}
\begin{split}
    S_k(t) 
&    =\left[ \frac{1}{2} + |\beta_k|^2 + |\alpha_k \beta_k|  \, \cos(  \theta_k + 2 \omega_k  t)\right](1 + 2 N_k^{\text{in}}). \\
\end{split}
\label{eq:ThermalSpectrum} 
\end{equation}
The term $1/2$ corresponds to vacuum fluctuations, the term $|\beta_k|^2$ describes the populations of a mode $k$ and the oscillating term contains the contributions from coherences. 
The Bogoliubov coefficients $\alpha_k$ and $\beta_k$, as well as the phase $\theta_k = \text{Arg}(\alpha_k\beta_k)$, depend on the expansion history and are calculated by relating the quantum states before and after the expansion \cite{BECPaper2022,CosmologyPaper2022}. The initial distribution is taken to be of thermal form
\begin{equation}
    N^{\text{in}}_k (T) = \frac{1}{e^{\hbar \omega_k/(k_\text{B} T)}-1}\;.
    \label{eq:Thermal}
\end{equation}
The experimentally extracted quantities are related to the spectrum by $A_k = |\alpha_k \beta_k|(1 + 2 N_k^{\text{in}})$ and ${\vartheta_k = \theta_k + 2 \omega_k \Delta t}$, with $\Delta t$ the duration of the ramp.

\bibliography{references.bib}

\newpage

\vfill
\renewcommand{\figurename}{\textbf{\textsf{Extended Data Figure|}}}
\renewcommand{\thefigure}{1}

\begin{figure*}
\centering
\includegraphics[width=0.9\textwidth]{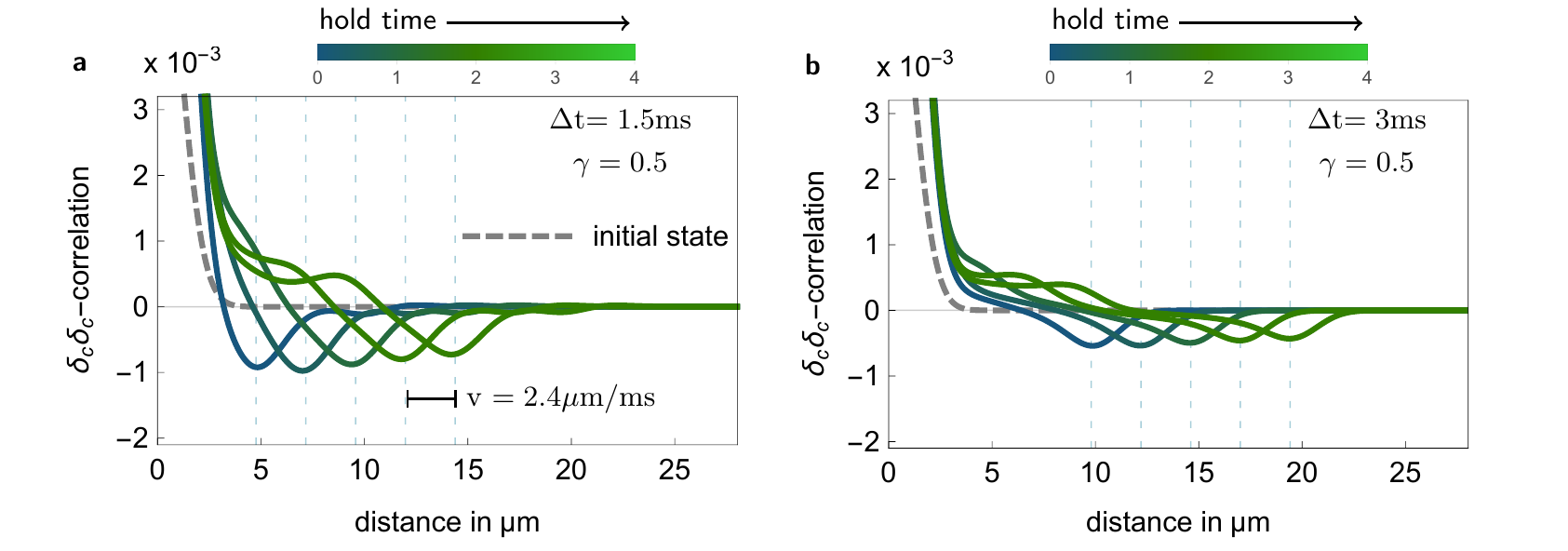}
\caption{\textsf{\textbf{Theoretical prediction for density-contrast correlation functions in real space.} The initial temperature is taken to be $40 \,\mathrm{n K}$, the final speed of sound is $1.2\, \mathrm{\mu m /m s }$ and the ramp goes from $a_s = 350\; \mathrm{a_\text{B}}$ to $a_s = 50\; \mathrm{a_\text{B}}$ with scale factor $a(t) \propto t ^\gamma$ for a decelerated expansion with $\gamma = 0.5$ and a duration $\Delta \mathrm{t}= 1.5 \, \mathrm{ms}$ (left) and $\Delta \mathrm{t}= 3.0 \, \mathrm{ms}$ (right). The fields are convoluted with a Gaussian of $\sigma = 0.8 \,\mathrm{\mu m}$ corresponding to the optical resolution of the experiment. Different colours correspond to different hold times. }}
\label{fig:extended_spatialcorrelation}
\end{figure*}

\end{document}